\documentclass[12pt]{article}
\pdfoutput=1

\usepackage{amsmath}
\usepackage{amssymb}
\usepackage{epsfig}
\usepackage{graphicx}
\usepackage{cite}
\usepackage{multirow}
\usepackage{longtable}
\usepackage{lscape}
\usepackage{bbm}
\usepackage{dcolumn}
\usepackage{rotating}

\allowdisplaybreaks[4] 

\newcommand{\capdef}{}
\newcommand{\mycaption}[2][\capdef]{\renewcommand{\capdef}{#2}%
        \caption[#1]{{\footnotesize #2}}}
\makeatletter
\renewcommand{\fnum@table}{\textbf{\tablename~\thetable}}
\renewcommand{\fnum@figure}{\textbf{\figurename~\thefigure}}
\makeatother

\newcounter{myenumi}

\renewcommand{\themyenumi}{\roman{myenumi}}
{\end{list}}

\newlength{\myem}
\newcounter{mysubequation}[equation]

\makeatletter
\renewcommand{\section}{\@startsection{section}{1}{0em}{-\baselineskip}%
{\baselineskip}{\normalfont\large\bfseries}}
\renewcommand{\subsection}%
{\@startsection{subsection}{2}{0em}{-0.7\baselineskip}%
{0.7\baselineskip}{\normalfont\bfseries}}
\makeatother

\newcommand{\bi}{\begin{itemize}}
\newcommand{\ei}{\end{itemize}}

\newcommand{\be}{\begin{equation}}
\newcommand{\ee}{\end{equation}}
\newcommand{\bea}{\begin{eqnarray}}
\newcommand{\eea}{\end{eqnarray}}

\newcommand{\ldm}{\Delta m_{31}^2}
\newcommand{\Ldm}{\Delta m_{41}^2}

\newcommand{\ie}{{\it i.e.}}

\newcommand{\eg}{{\it e.g.}}

\newcommand{\eq}{Eq.}

\newcommand{\fig}{Fig.}

\newcommand{\Ref}{Ref.}
\newcommand{\Refs}{Refs.}
\newcommand{\Sec}{Sec.}

\newcommand{\equ}[1]{\eq~(\ref{equ:#1})}
\newcommand{\figu}[1]{\fig~\ref{fig:#1}}

\begin{document}

\begin{titlepage}

\renewcommand{\thefootnote}{\alph{footnote}}

\vspace*{-3.cm}
\begin{flushright}
IDS-NF-034 \\
EURONU-WP6-12-48
\end{flushright}

\vspace*{0.5cm}

\renewcommand{\thefootnote}{\fnsymbol{footnote}}
\setcounter{footnote}{-1}

{\begin{center}
{\large\bf
 Optimization of a Very Low Energy Neutrino Factory for the Disappearance Into Sterile Neutrinos
}

\end{center}}

\renewcommand{\thefootnote}{\alph{footnote}}

\vspace*{.8cm}
\vspace*{.3cm}
{\begin{center} {\large{\sc
                Walter~Winter\footnote[1]{\makebox[1.cm]{Email:}
                winter@physik.uni-wuerzburg.de}
                }}
\end{center}}
\vspace*{0cm}
{\it
\begin{center}

\footnotemark[1]
       Institut f{\"u}r Theoretische Physik und Astrophysik, \\ Universit{\"a}t W{\"u}rzburg,
       97074 W{\"u}rzburg, Germany

\end{center}}

\vspace*{1.5cm}


{\Large \bf
\begin{center} Abstract \end{center}  }

We discuss short-baseline electron and muon neutrino disappearance searches into sterile neutrinos at a Very Low Energy Neutrino Factory (VLENF) with a muon energy between about two and four GeV. A lesson learned from reactor experiments, such as Double Chooz and Daya Bay, is to use near and far detectors with identical technologies to reduce the systematical errors. We therefore derive the physics results from a combined near-far detector fit and illustrate that uncertainties on cross sections $\times$ efficiencies can be eliminated in a self-consistent way. We also include the geometry of the setup, \ie, the extension of the decay straight and the muon decay kinematics relevant at the near detector, and we demonstrate that these affect the sensitivities for $\Delta m^2 \gtrsim 30 \, \mathrm{eV}^2$, where oscillations take place already in the near detector. Compared to appearance searches, we find that the sensitivity depends on the locations of both detectors and the muon energy, where the near detector should be as close as possible to the source, and the far detector at about 500 to 800~m. In order to  exclude  the currently preferred parameter region, at least $10^{19}$ useful muon decays per polarity are needed for $E_\mu=2 \, \mathrm{GeV}$, or, alternatively, a higher muon energy can be used. 

\vspace*{.5cm}

\end{titlepage}

\newpage

\renewcommand{\thefootnote}{\arabic{footnote}}
\setcounter{footnote}{0}

\section{Introduction}

The test of three-flavor neutrino oscillations in solar, atmospheric, long-baseline, and reactor experiments has been so far very successful, where a non-zero mixing angle $\theta_{13}$ has been recently established by the Daya Bay and RENO reactor experiments above the $5 \sigma$ confidence level~\cite{An:2012eh,Ahn:2012nd}. On the other hand, neutrino oscillations at short baselines, \ie, $L \ll E/\ldm$ where atmospheric oscillations have not yet developed, face a tension between the strong constraints from many short-baseline experiments, and several observed anomalies which may be described by eV-scale sterile neutrinos. In particular, short-baseline electron neutrino disappearance may cause an anomaly  identified in Gallium experiments~\cite{Acero:2007su},  electron antineutrino neutrino disappearance may lead to lower than predicted reactor antineutrino fluxes~\cite{Mention:2011rk,Huber:2011wv}, and electron antineutrino appearance may be driven by sterile neutrinos in the LSND~\cite{Aguilar:2001ty} and MiniBooNE~\cite{AguilarArevalo:2010wv} experiments. In the simplest models, one would add one extra sterile generation to fit these data by a $3+1$ model, separated by $\Delta m^2 \sim 1 \, \mathrm{eV}^2$, see, \eg, \Ref~\cite{Giunti:2011hn}. However, it turns out that the tension between MiniBooNE neutrino and antineutrino data (CP violation cannot be described by this model), and the tension between appearance and disappearance data basically rule out this model, see, \eg, \Ref~\cite{Maltoni:2007zf}. Therefore, $3+2$ models have been more recently used~\cite{Kopp:2011qd}, which allow for CP violation and include additional degrees of freedom. In any of the above models, crucial information comes from the disappearance channels, which may be the ``cleanest'' channels to measure the oscillation parameters (see \Sec~\ref{sec:pheno} for details). In addition, electron neutrino and antineutrino disappearance searches are needed to directly test the Gallium and reactor anomalies, respectively. 

Various new experiments have been proposed to test short-baseline neutrino oscillations, see Appendix~A of \Ref~\cite{SNAC} for a recent summary of alternatives. In this study, we focus on a very low energy neutrino factory (VLENF), which is a neutrino factory with a low muon energy of about two to four GeV which does not require muon cooling or muon acceleration~\cite{Tunnell:2011ya}, and could be the first phase of a staged neutrino factory program. Compared to muon neutrino appearance, discussed in \Ref~\cite{Tunnell:2011ya}, we discuss the electron and muon neutrino (antineutrino) disappearance channels. These are qualitatively different from the appearance channels because of different systematics: while the appearance channels are limited by (charge mis-identification and neutral current) backgrounds, the disappearance channels suffer from unknown cross sections $\times$ efficiencies. Therefore, similar to reactor experiments such as Double Chooz or Daya Bay, near and far detectors have been proposed for the high energy neutrino factory to control the systematical errors~\cite{Giunti:2009en}.
In addition, near detectors very close to the muon storage ring experience geometry effects from the extension of the decay straight and beam divergence~\cite{Tang:2009na}. In \Ref~\cite{Giunti:2009en} the geometry effects from
the decay straight have been taken into account, which lead to a smearing of the oscillation probabilities, whereas the detectors were assumed to be far enough away from the source (or small enough) not to experience any detector geometry effects (far distance limit). However, for the substantially lower muon energy of the VLENF, the beam will be larger from the muon decay kinematics only, which means that the detector geometry has to be taken into account. In addition, the near and far detectors are supposed to be as similar as possible, which is difficult to reconcile with different detector diameters. Therefore, we will simultaneously integrate over straight geometry and detector surface area in this study, and show the impact of these effects. We discuss electron neutrino disappearance for most of this study, since it is directly relevant to test some of the anomalies, and we point out the differences for muon neutrino disappearance at the end.

This study is organized as follows: we recapitulate the phenomenology of sterile neutrino disappearance searches in \Sec~\ref{sec:pheno}. Then in \Sec~\ref{sec:sim}, we introduce our setup, systematics treatment, and geometry treatment. In \Sec~\ref{sec:geosys}, we illustrate the impact of the beam and detector geometry, and of the systematics assumptions. Then in \Sec~\ref{sec:twobase}, we perform a two-baseline optimization of near and far detectors. We show our main results for electron and muon neutrino disappearance in \Sec~\ref{sec:results}, and we conclude in \Sec~\ref{sec:summary}.

\section{Sterile neutrino phenomenology}
\label{sec:pheno}

In most experiments, the disappearance or appearance of neutrinos at short distances is described in the two-flavor limit
\begin{eqnarray}
P_{\alpha \alpha} & = & 1 - \sin^2 (2 \theta_{\mathrm{eff}}) \sin^2 \Delta  \, , \label{equ:disapp} \\
P_{\alpha \beta} & = & \sin^2 (2 \theta_{\mathrm{eff}}) \sin^2 \Delta  \, , \label{equ:app}
\end{eqnarray}
where $\Delta \equiv \Delta m^2 L/(4 E)$ and $\theta_{\mathrm{eff}}$ is an effective mixing angle. However, this description cannot be used for a self-consistent description of a multi-channel experiment.
In order to demonstrate that, consider the simplest possible case, a $3+1$ framework with one extra sterile neutrino. Although this case is basically excluded, it is useful two illustrate some of the considerations to be taken into account here which also apply to more sophisticated models.
The parameterization-independent probabilities in the limit $\Delta_{41}  \gg \Delta_{31} \simeq 0$ ($\Delta_{ij} \equiv \Delta m_{ij}^2 L/(4E))$ can be written as 
\begin{eqnarray}
P_{ee} & =& 1- 4 | U_{e4}|^2 (1-|U_{e 4}|^2) \sin ^2 \Delta_{41} \, , \label{equ:pee} \\
P_{\mu\mu}& =& 1- 4 | U_{\mu 4}|^2 (1-|U_{\mu 4}|^2) \sin ^2 \Delta_{41} \, , \label{equ:pmm} \\
P_{e\mu} = P_{\mu e} & =& 4 | U_{e 4}|^2 |U_{\mu 4}|^2 \sin ^2 \Delta_{41} \, , \label{equ:pem}
\end{eqnarray}
where we show the most interesting channels for the Neutrino Factory. From this simple model, it is immediately clear that LSND-motivated electron or  muon neutrino appearance, which requires $| U_{e 4}| >0 $ and $|U_{\mu 4}| > 0$, must be accompanied by electron {\em and} muon neutrino disappearance, and that the disappearance searches provide important and strict constraints on theoretical models.  Especially, the disappearance probabilities are proportional to $|U_{\alpha 4}|^2$, the appearance probabilities to $|U_{\alpha 4}|^4$, \ie, the appearance probabilities are suppressed by two more powers of the new mixing angles.
In a specific parameterization, electron and muon disappearance can be described by the two-flavor limit \equ{disapp} by different mixing angles (such as $\theta_{14}$ and $\theta_{24}$, respectively), whereas the appearance probability in \equ{app} is proportional to the product of these mixing angles; see \eg\ \Ref~\cite{Meloni:2010zr} for a direct comparison. This means that \equ{disapp} or \equ{app} can be used independently to effectively describe an individual disappearance or appearance channel. If, however, the information from different channels is to be combined, there will be a model-dependent (but well defined) interplay among the channels.

Now what are the consequences for the Neutrino Factory? For electron or muon neutrino appearance, the main limitations are charge mis-identification\footnote{For instance, for neutrino production by $\mu^+ \rightarrow e^+ + \nu_e + \bar\nu_\mu$, one has to distinguish a few $\nu_e \rightarrow \nu_\mu$ charged current events from the dominant $\bar \nu_\mu \rightarrow \bar \nu_\mu$ event rate by charge-identification of the produced muons (anti-muons) by means of a magnetic field.} and neutral current backgrounds. At least for small mixing angles, other systematics, such as the flux and cross section uncertainties, are less relevant. That is quite fortunate, since it would be probably hard to quantify this systematics. Consider, for instance, a near detector to measure the cross sections for muon neutrinos.
For large enough $\Ldm$, oscillations may already take place in the near detector. Unless it is clearly defined how the theoretical interplay between muon neutrino disappearance at the near detector and muon neutrino  appearance at the far detector works, one cannot disentangle the cross sections from oscillation physics in that case. Thus, an effective two-flavor description is not sufficient to obtain reliable information on this systematics.

For the disappearance channels, the situation is very different. Here the electron and muon neutrino disappearance can be described by \equ{disapp} in the effective two-flavor limit with different effective mixing angles. Because small deviations from unity are to be measured, a lesson learned from reactor experiments~\cite{Minakata:2002jv,Huber:2003pm}, such as Double Chooz or Daya Bay, has been to use (more or less) identical near and far detectors to cancel systematical errors. Compared to reactor experiments, where the flux is the unknown, the cross section $\times$ efficiency uncertainties are the dominant systematics to be canceled by the near detector. From the oscillation framework point of view, since both near and far detectors measure the same flavors, the oscillation probabilities in both detectors can be described by the same probabilities. In the $3+1$ model, one can simply use \equ{disapp} for $\nu_e$ or $\nu_\mu$ disappearance. In a $3+N$ model, the translation into the model parameters is more complicated. Nevertheless, it is clear that any detected difference between near and far rates (or neutrino and antineutrino rates) must be due to oscillations, and cannot come from unknown cross sections or efficiencies.
Very interestingly, the near and far detectors may change their roles as a function of $\Ldm$. While for small $\Ldm$ the far detector measures the oscillations and the near detector the normalization, for large $\Ldm$ the near detector measures the oscillations  and the far detector, where the oscillation averages out, the normalization~\cite{Giunti:2009en}. We choose an effective two-flavor framework in the following to quantify the performance, and we assume CPT invariance for the sake of simplicity (the disappearance channels are always CP invariant). Note that nevertheless CPT invariance tests of the disappearance channels are well motivated, see \Ref~\cite{Giunti:2009en} for the Neutrino Factory. We do not perform a combined fit of appearance and disappearance channels, or electron and muon disappearance channels, since such a fit can only be performed for a specific model. Note, however, that the final experiment can of course be used to test more complicated scenarios.

\section{Setup and simulation techniques}
\label{sec:sim}

\begin{figure}
\begin{center}
\includegraphics[angle=-90,width=\textwidth]{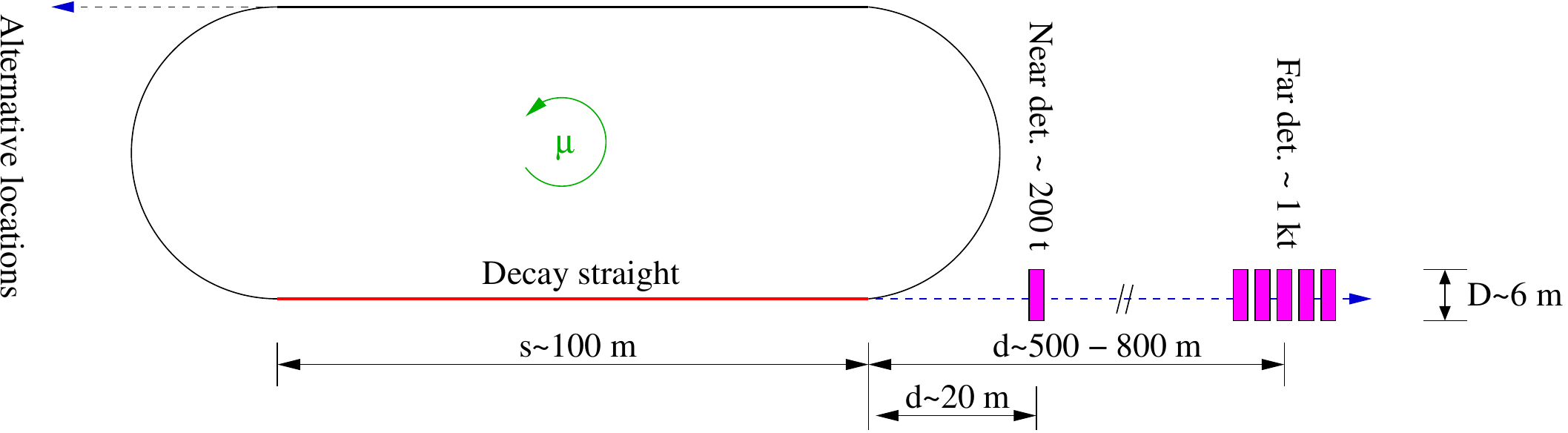}
\end{center}
\caption{\label{fig:ring} Possible geometry of the VLENF near-far setup for $E_\mu = 2 \, \mathrm{GeV}$ (not to scale). }
\end{figure}

For the simulation, we choose a near-far detector setup with a near detector of $200 \, \mathrm{t}$ fiducial mass $\times$ efficiency, and a far detector of $1 \, \mathrm{kt}$ fiducial mass $\times$ efficiency, see \figu{ring} for a possible geometry. Since the detector geometry will be important for close distances, we assume cylindrical shapes with diameters of $6 \, \mathrm{m}$ (perpendicular to the beam axis). We test two different muon energies: $E_\mu=2 \, \mathrm{GeV}$ and $E_\mu=4 \, \mathrm{GeV}$. The decay straight length $s$ is assumed to be $100 \, \mathrm{m}$ for $E_\mu=2 \, \mathrm{GeV}$, and $200 \, \mathrm{m}$ for $E_\mu=4 \, \mathrm{GeV}$. For very short baselines and line neutrino sources, the baseline $L$ is ill-defined~\cite{Tang:2009na}. Therefore, the detector locations are specified by the distance $d$ to the end of the decay straight, \ie,  $d+s \le L \le d$. For the integrated luminosity, we use $10^{19}$ useful muon decays per polarity, and  $10^{18}$ useful muon decays in some cases where explicitely specified.
For the detection threshold, we use $0.25 \, \mathrm{GeV}$ and $0.5 \, \mathrm{GeV}$ for $E_\mu=2 \, \mathrm{GeV}$ and $E_\mu=4 \, \mathrm{GeV}$, respectively. For the energy resolution of the detectors, we use $\Delta E/\mathrm{GeV} = 0.1 \, \sqrt{E/\mathrm{GeV}}$ for $\nu_e$ ($\bar \nu_e$) disappearance, and $\Delta E/\mathrm{GeV} = 0.05 \, \sqrt{E/\mathrm{GeV}}$ for $\nu_\mu$ ($\bar \nu_\mu$) disappearance, which is motivated by a totally active scintillator or a liquid argon detector. Neutral current backgrounds are included at the level of $10^{-4}$, which have, however, only a very small effect. Note that a magnetization of the detector for the disappearance searches may or may not be necessary, depending on the underlying physics model, \ie, what happens to the other flavor. For instance, in the $3+1$ model discussed above, one may have large $\nu_\mu$ disappearance driven by non-zero $|U_{\mu4}|$, see \equ{pmm}, whereas $\bar \nu_\mu$ appearance vanishes at the same time  for $|U_{e4}|=0$, see \equ{pem}. Since we only use the disappearance channels, we assume that charge mis-identification is either under control by a magnetic field, or suppressed by the underlying physics.
 Note that the chosen parameters are consistent with the currently discussed ones of the VLENF study group~\cite{Bross}; see also \Ref~\cite{Tunnell:2011ya}.

For the geometric treatment of the neutrino line source and detector geometry, we follow \Refs~\cite{Tang:2009na,Giunti:2009en}, and for the simulation, we use the GLoBES (General Long Baseline Experiment Simulator) software~\cite{Huber:2004ka,Huber:2007ji}. GLoBES (up to version 3.1) assumes that the detectors are far enough away from the source to treat the source as point source, and that the detectors are small compared to the beam divergence. If we start from the differential event rate from a point source $dN_{\mathrm{PS}}/dE$ without oscillations, as the one used in GLoBES, we can take into account the extension of the straight and the detector and the effect of oscillations by an averaged event rate
\begin{equation}
\frac{dN_{\mathrm{avg}}}{dE} = \frac{1}{s} \int\limits_{d}^{d+s}\frac{dN}{dE} dL = \frac{1}{s} \int\limits_{d}^{d+s}\frac{dN_{\mathrm{PS}}(L,E)}{dE} \, \varepsilon_\alpha(L,E) \, P_{\alpha \alpha}(L,E) dL \, .
\end{equation}
 Here $\varepsilon_\alpha(L,E)=A_{\mathrm{eff}, \alpha}/A_{\mathrm{Det}}$
parameterizes the integration over the detector geometry for a fixed baseline $L$, energy $E$, and flavor $\nu_\alpha$, where $A_{\mathrm{eff}, \alpha}(L, E)$ is the (energy and flavor dependent) effective area of the detector, and  $A_{\mathrm{Det}}$ is the actual surface area. Note that the oscillation probability $P_{\alpha \alpha}$ appears inside the integral, since different parts of the decay straight contribute differently. In addition, note that it is assumed that the differential muon decay rate per (straight) length is equal along the decay straight. Since $dN_{\mathrm{PS}}/dE \propto 1/L^2$,
we can re-write this as
\begin{equation}
\frac{dN_{\mathrm{avg}}}{dE} = \frac{dN_{\mathrm{PS}}(L_{\mathrm{eff}},E)}{dE} \frac{L_{\mathrm{eff}}^2}{s} \int\limits_{d}^{d+s} \frac{ \varepsilon_\alpha(L,E)}{L^2} \, P_{\alpha \alpha}(L,E) dL  =  \frac{dN_{\mathrm{PS}}(L_{\mathrm{eff}},E)}{dE}
 \, \hat P(E) \label{equ:eavg}
\end{equation}
with the average efficiency ratio times probability
\begin{equation}
  \hat P(E)  \equiv \frac{L_{\mathrm{eff}}^2}{s} \int\limits_{d}^{d+s} \frac{ \varepsilon_\alpha(L,E)}{L^2} \, P_{\alpha \alpha}(L,E) dL
\label{equ:peff}
\end{equation}
and the effective baseline
\begin{equation}
L_{\mathrm{eff}}=\sqrt{d (d+s) } \, , \label{equ:leff}
\end{equation}
which is defined such that $\hat P(E)=1$ for $\varepsilon_\alpha(L,E) \equiv 1$ and $P_{\alpha \alpha}(L,E)\equiv 1$. As a consequence, 
one would use the usual detector definitions in GLoBES with the effective baseline $L_{\mathrm{eff}}$, which are to be corrected by \equ{peff}. We compute \equ{peff} directly in a user-defined probability engine in GLoBES, including both neutrino source and detector geometry. For the sake of simplicity, we assume that the (machine-dependent) beam divergence is smaller the the beam spread given by the muon decay kinematics. In this ideal case, one can easily compute $\varepsilon_\alpha(L,E)$ independent of machine-dependent effects more or less analytically~\cite{Tang:2009na}. We will demonstrate that there is a substantial effect coming from the extension of the beam compared to the detector, whereas machine-dependent effects may lead to additional smearing if the divergence is not under control. Note that the intrinsic effect of the muon decay kinematics cannot be removed, even in an ideal machine. 

For the systematics treatment, we follow the reactor experiments with two (or more) detector, rather than the usual neutrino factory description; see \Refs~\cite{Huber:2006vr,Giunti:2009en} for details. From the reactor experiments, we known that  short baseline electron neutrino disappearance is
mostly affected by the signal normalization uncertainty (see, \eg, \Refs~\cite{Huber:2003pm,Huber:2006vr} 
for reactor experiments). Here, compared to
the reactor experiments, our signal normalization error is not dominated by the flux, which
may be known at the level of 0.1\% using various mean monitoring devices~\cite{Abe:2007bi},
but the knowledge of the cross sections $\times$ efficiencies. Because our neutrino energies span the
cross section regimes from quasi-elastic scattering, over resonant pion production, 
to deep inelastic scattering, it is difficult to estimate the
degree the cross sections will be known at the time of the measurement. The efficiencies  depend on the detection processes and detector properties, which means that their uncertainties are also difficult to pin down. For reactor experiments,
on the other hand, the inverse beta decay cross sections are well known, but the fluxes are relatively uncertain. Both classes of experiments have, however, in common that these uncertainties can be controlled by using detectors as identical as possible. In our case, where the near and far detector masses are different, the far detector may consist of five modules identical to the near detector, as it is illustrated in \figu{ring}. We neglect the extension of the detector along the beam axis, since it is expected to me much smaller than the extension of the decay straight. However, in practice, the location of the interaction vertex can be measured to some degree. 

We adopt the most conservative case for systematics, which is that the cross sections $\times$ efficiencies ($\times$ flux) are fully uncorrelated among the bins, but fully correlated between the near and far detectors.\footnote{We use 17~bins with bin widths following the energy resolution of the detector. To avoid aliasing effects, we use the built-in filter of GLoBES on the oscillation probability.} This assumption is conservative because is corresponds to  cross sections $\times$ efficiencies with an unknown shape error, where the shape is to be measured by the near detector. Unless noted otherwise, we assume that cross sections $\times$ efficiencies are only known to the level of 10\% (within each bin), where even larger errors do not matter in the oscillation region.
In addition, we use a normalization error uncorrelated between near and far detectors, but fully correlated among the bins, which may come from a fiducial mass uncertainty. This error is, for reactor experiments, believed to be controlled below the percent level. We use $0.6\%$~\cite{Huber:2006vr}, whereas Daya Bay claim that they can  even do significantly better ($0.2\%$~\cite{An:2012eh}). Since it depends on the detector properties and detection interactions, a more conservative choice seems reasonable. Finally, the backgrounds are assumed to be known within 35\% and the energy calibration to 0.5\%, uncorrelated between the detectors. In the next section, we will discuss where these uncertainties are important.

\section{Impact of geometry and systematics}
\label{sec:geosys}

\begin{figure}[t!]
\begin{center}
\includegraphics[width=0.48\textwidth]{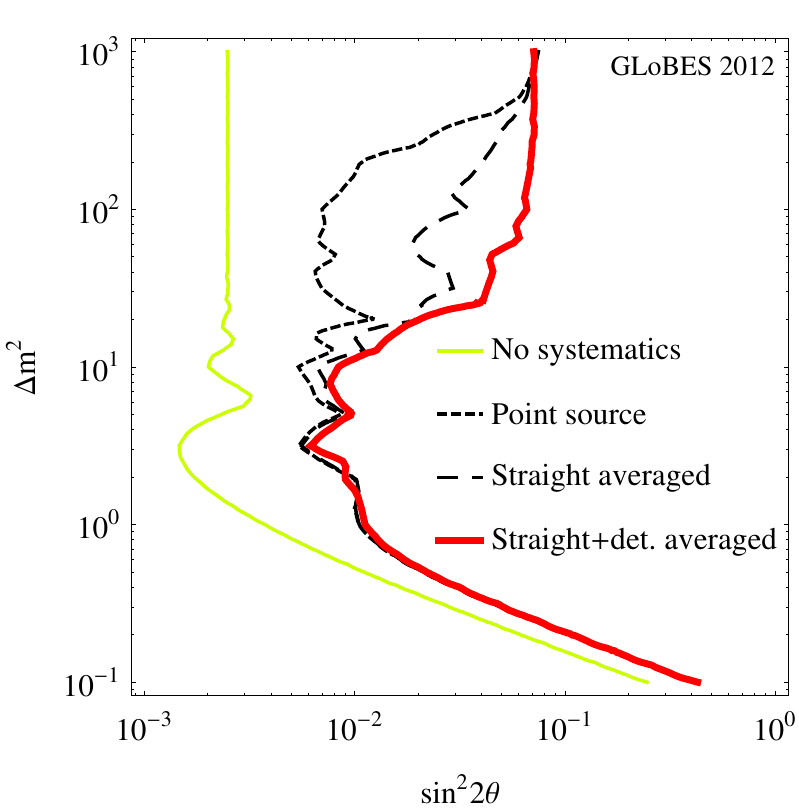} \hspace*{0.02\textwidth} \includegraphics[width=0.48\textwidth]{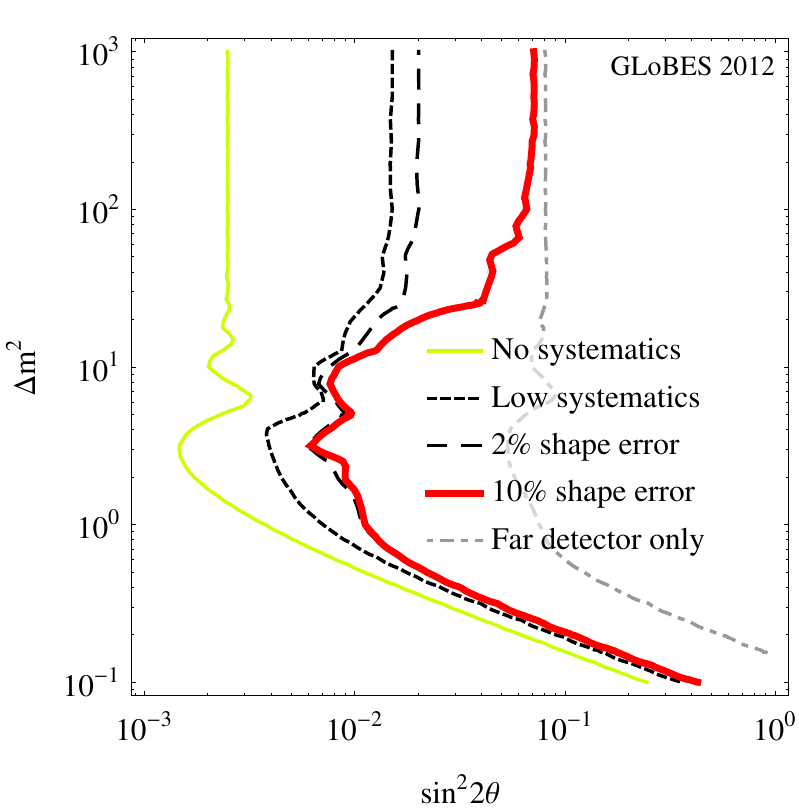}
\end{center}
\mycaption{\label{fig:sys} Exclusion region in $\sin^2 2 \theta$-$\Delta m^2$ (right hand sides of curves) for $\nu_e$ disappearance for different geometry assumptions (left panel) and systematics assumptions (right panel); see main text for details (90\% CL, 2 d.o.f.). The curves ``no systematics'' represents a single detector at $d=500 \, \mathrm{m}$ using statistics only, whereas the other curves correspond to near-far detector setups, where the red thick curves include (conservative) full systematics and geometry effects. Here $E_\mu=2 \, \mathrm{GeV}$, $10^{19}$ useful muon decays per polarity, $d_1=20 \, \mathrm{m}$ ($200 \, \mathrm{t}$) and $d_2 = 500 \, \mathrm{m}$ ($1 \, \mathrm{kt}$). }
\end{figure}

Let us now study the impact of geometry and systematics, where we use $E_\mu=2 \, \mathrm{GeV}$,  $d_1=20 \, \mathrm{m}$ ($200 \, \mathrm{t}$), and $d_2 = 500 \, \mathrm{m}$ ($1 \, \mathrm{kt}$). For the most of the following, we focus on $\nu_e$ ($\bar \nu_e$) disappearance, and discuss the differences to muon neutrino disappearance at the end. In \figu{sys}, left panel, 
the exclusion region in $\sin^2 2 \theta$ and $\Delta m^2$ (right hand sides of curves) is shown for different geometry assumptions. As we will demonstrate below, this setup is optimized for not too small values of $\Delta m^2$. The curve ``no systematics'' represents the statistics limit of the far detector only, without systematics or backgrounds. It is simulated in the point source and far distance approximations, which means that the baseline is computed with \equ{leff} and that $\varepsilon_\alpha(L,E)\equiv 1$ in \equ{peff}, respectively, as it is usually done for far detectors. The curve ``point source'' is obtained for the same geometry assumptions, but for a near-far detector simulation including full systematics. The first peak of the sensitivity at about $\Delta m^2 \simeq 3 \, \mathrm{eV}^2$ corresponds to the far detector, as it can be easily seen. In this case, the near detector measures the normalization, the far detector the oscillation.  The near detector sensitivity peaks at about $\Delta m^2 \simeq 13 \, \mathrm{eV}^2$, where, however, oscillations are still present in the far detector (see ``no systematics'' curve). Therefore, the optimal sensitivity is reached at somewhat larger values of $\Delta m^2$, where oscillations in the far detector average out and the cross sections $\times$ efficiencies are safely measured in the far detector. In this case, near and far detectors swap their roles. This swapping is also the reason why it is difficult to obtain reliable sensitivity predictions for effective far detector simulations only, especially for $\Delta m^2 \gg 1 \, \mathrm{eV}^2$, where oscillations take place in the near detector.
As the next step in \figu{sys}, left panel, we take into account the extension of the decay straight in the curve ``straight averaged'', which leads to some averaging in the region the near detector is most sensitive to. This case corresponds to using \equ{peff} with $\varepsilon_\alpha(L,E)\equiv 1$.
The final result, curve ``straight and detector averaged'', shows the additional impact of the detector geometry, \ie, the full \equ{peff}. In this case, the sensitivity for large $\Delta m^2$, coming from the near detector, almost vanishes. The reason is that the detected spectrum effectively peaks at lower energies due to the muon decay kinematics, which the near detector (but not the far detector) is sensitive to.  For very large $\Delta m^2 \gg 100 \, \mathrm{eV}^2$, the oscillations average out in both near and far detectors, and the sensitivity is given by the externally imposed error of 10\%.

We have also tested if one can reduce the effect of the straight averaging by the use of two beam current transformers (BCTs), one before and one after the straight. One may assume that the number of muon decays along the straight, which is proportional to the difference between the two beam currents,  is exponentially distributed along the straight. However, already at $E_\mu = 2 \, \mathrm{GeV}$, the mean lifetime (decay length) of the muon is about 6~km, which means that the effect is small and not visible in the sensitivities.  

In the right panel of \figu{sys}, we show the impact of different systematics assumptions, where again the ``no systematics'' and full systematics (``10\% shape error'') curves are used for reference. First of all, we reduced the systematical errors one by one. As expected, a smaller shape error (curve ``2\% shape error'') improves the sensitivity for large $\Delta m^2$, where the oscillations in near and far detectors are averaged out and this systematics dominates. None of the other systematics has an effect on the sensitivity if improved separately. Only if all of the systematical errors can be controlled better, the curve ``low systematics'' is obtained.\footnote{Low syst.: Fiducial volume/normalization error 0.1\%, shape error 2\%, calibration error 0.1\%, background error 10\%.} Since it is difficult to say how realistic this case is, we rely on our standard (more conservative) values in the following, unless explicitely stated otherwise. In \figu{sys}, we also show the result for the far detector only (gray dashed-dotted curve), from which one can see  that the near detector is necessary to maintain the sensitivity for low $\Delta m^2 \simeq \mathrm{eV}^2$ in the presence of the systematical errors.
 We include the decay straight and detector geometries in the following calculations, as well as the near detector, where the full systematics curve (``10\% shape error'') represents our standard values.

\section{Two-baseline optimization}
\label{sec:twobase}

\begin{figure}[t!]
\begin{center}
\includegraphics[width=0.44\textwidth]{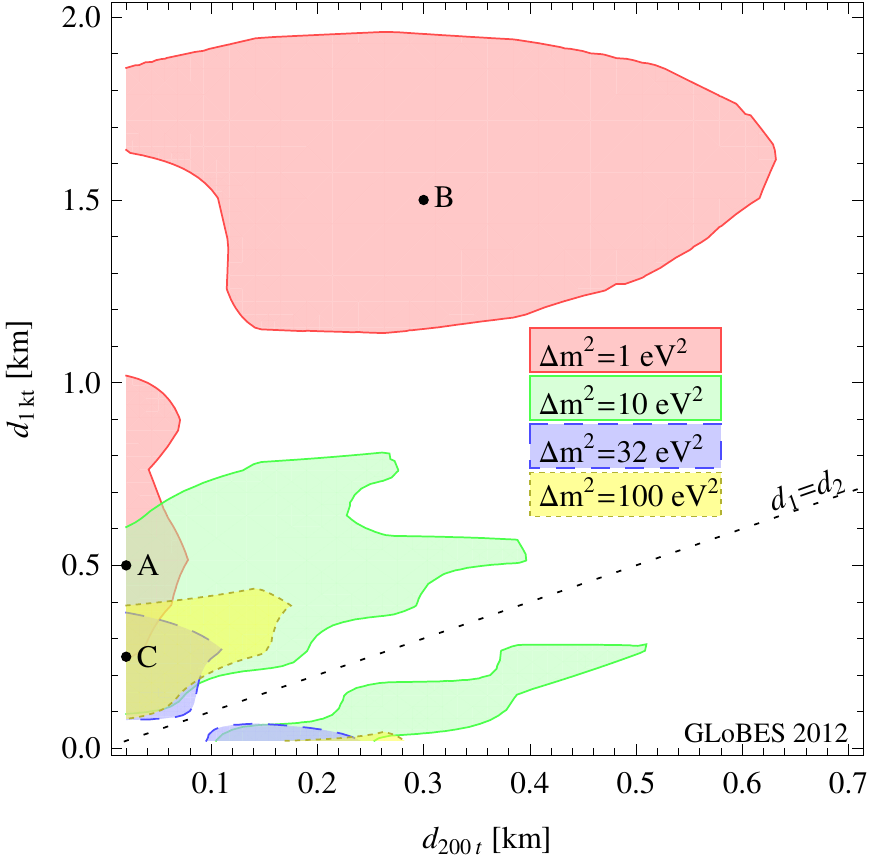} \hspace{0.03\textwidth}
\includegraphics[width=0.45\textwidth]{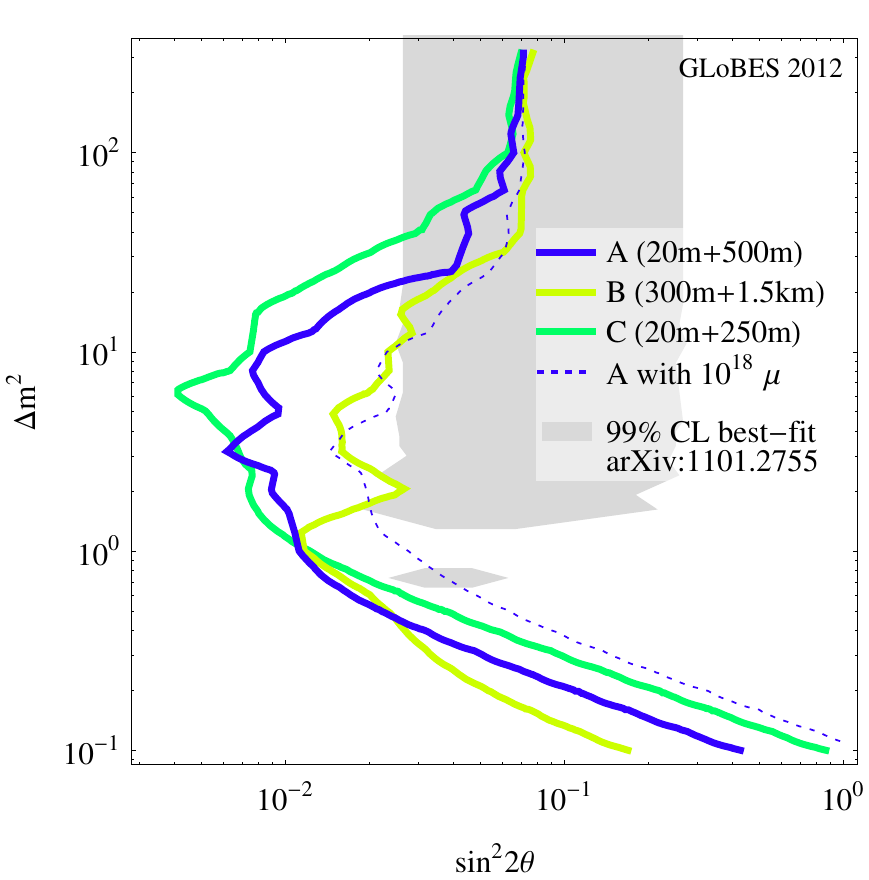} 
\end{center}
\mycaption{\label{fig:opt2} Left panel: Optimal detector distances for $\nu_e$ disappearance, $E_\mu=2 \, \mathrm{GeV}$, $10^{19}$ useful muon decays per polarity. The contours correspond to sensitivities in $\mathrm{log}_{10} (\sin^2 2 \theta)$ down to -1.9 ($\Delta m^2 = 1 \, \mathrm{eV}^2$), -1.95 ($\Delta m^2 = 10 \, \mathrm{eV}^2$), -1.5  ($\Delta m^2 = 32 \, \mathrm{eV}^2$), -1.2 ($\Delta m^2 = 100 \, \mathrm{eV}^2$)  for the depicted values of $\Delta m^2$ (90\% CL, 2 d.o.f.). Right panel: Sensitivities in $\sin^2 2 \theta$-$\Delta m^2$ for the ``optimal'' setups marked in the left panel. For comparison, setup~A is shown with $10^{18}$ useful muon decays per polarity, and the best-fit region from \Ref~\cite{Mention:2011rk} (Fig.~8) is shown as gray-shaded region (99\% CL). }
\end{figure}

Whereas the muon energy is limited by other constraints, such as accelerator and storage ring, the detector locations of the two detectors can be almost freely chosen. We therefore show in \figu{opt2}, left panel,  the two-baseline optimization for $\nu_e$ disappearance. The different regions correspond to optimal detector locations for the depicted values of $\Delta m^2$, where the respective reaches in $\sin^2 2 \theta$ are given in the figure caption. For $\Delta m^2 = 1 \, \mathrm{eV}^2$, two optimal regions are found, where two central choices are marked by points~A and~B. Point~A is within a region where the near detector is as close as possible to the source, and the far detector in a distance between about $200 \, \mathrm{m}$ and $1 \, \mathrm{km}$. Point~B corresponds to a longer far detector baseline which helps for small values of $\Delta m^2$. However, a somewhat farther near detector distance is preferred, where the near detector also adds to the sensitivity directly. The larger $\Delta m^2$ is, the smaller distances for the far detector are preferred, where in all cases the near detector may be as close as possible to the source. Point~C is representative for a region with better sensitivity for $\Delta m^2 \gtrsim 10 \, \mathrm{eV}^2$. The asymmetry in \figu{opt2} (left panel) with respect to the symmetry axis $d_1=d_2$ comes from the different detector masses. For equal detector masses, we do not find a qualitative differences apart from the plot becoming symmetric with respect to this axis.

The results in the $\Delta m^2$-$\sin^2 2 \theta$-plane are shown for the marked setups in \figu{opt2}, right panel. Setup~B has the best sensitivity for small values of $\Delta m^2$, but for larger $\Delta m^2$ it is not optimal. Setup~C is best for  $\Delta m^2 \gtrsim 1 \, \mathrm{eV}^2$, as expected, whereas setup~A is a good compromise between the small and large $\Delta m^2$ sensitivities. Therefore, we have chosen it for reference.  In order to check if the chosen setup match the needs for $\nu_e$ disappearance, we show the best-fit region from \Ref~\cite{Mention:2011rk} (Fig.~8) as gray-shaded region for reference (99\% CL). One can easily read off the figure that all setups can exclude this best-fit region very well, with setup~C actually covering the largest part. Note again that the very large $\Delta m^2$ coverage is limited by the external knowledge on cross sections $\times$ efficiencies for the VLENF, whereas the flatness of the best-fit region  for large $\Delta m^2$ simply means that the reactor experiments cannot resolve the oscillations (necessary to exclude this part). Therefore, one should probably not overemphasize this part. For reference, we also show in \figu{opt2}, right panel, the curve for point~A with 10$^{18}$ useful muon decays (per polarity) only, as different luminosities are currently being discussed.  One can clearly read off the figure, that the statistics is not sufficient to fully exclude the interesting part of the best-fit region at lower values of $\Delta m^2$.

\begin{figure}[t!]
\begin{center}
\includegraphics[width=0.44\textwidth]{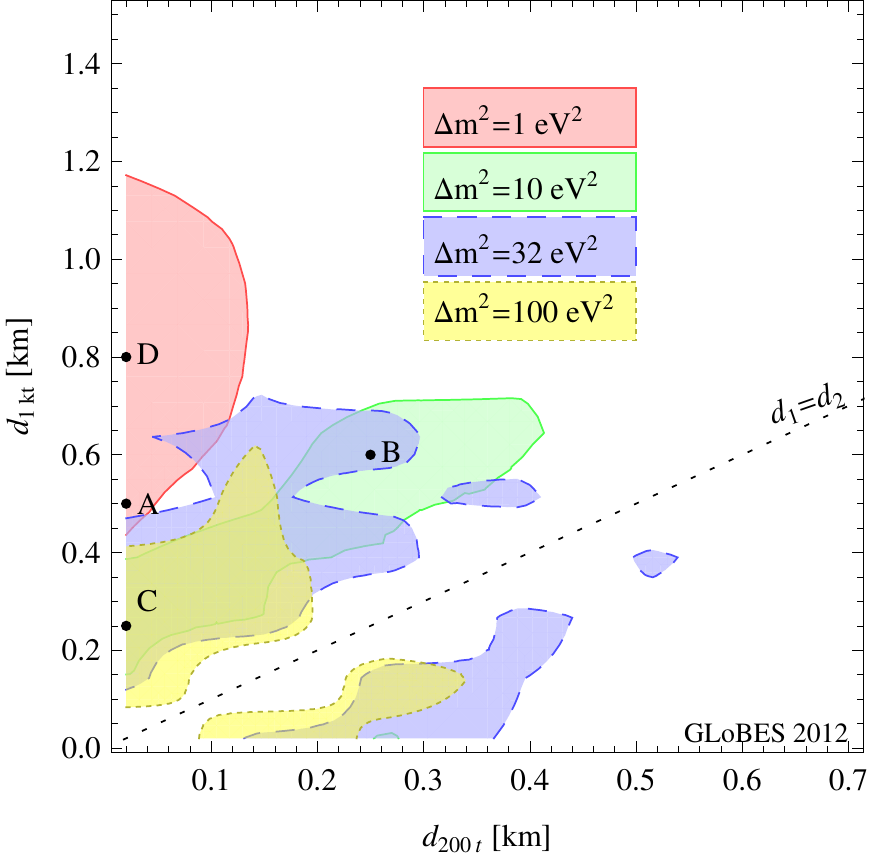} \hspace{0.03\textwidth}
\includegraphics[width=0.45\textwidth]{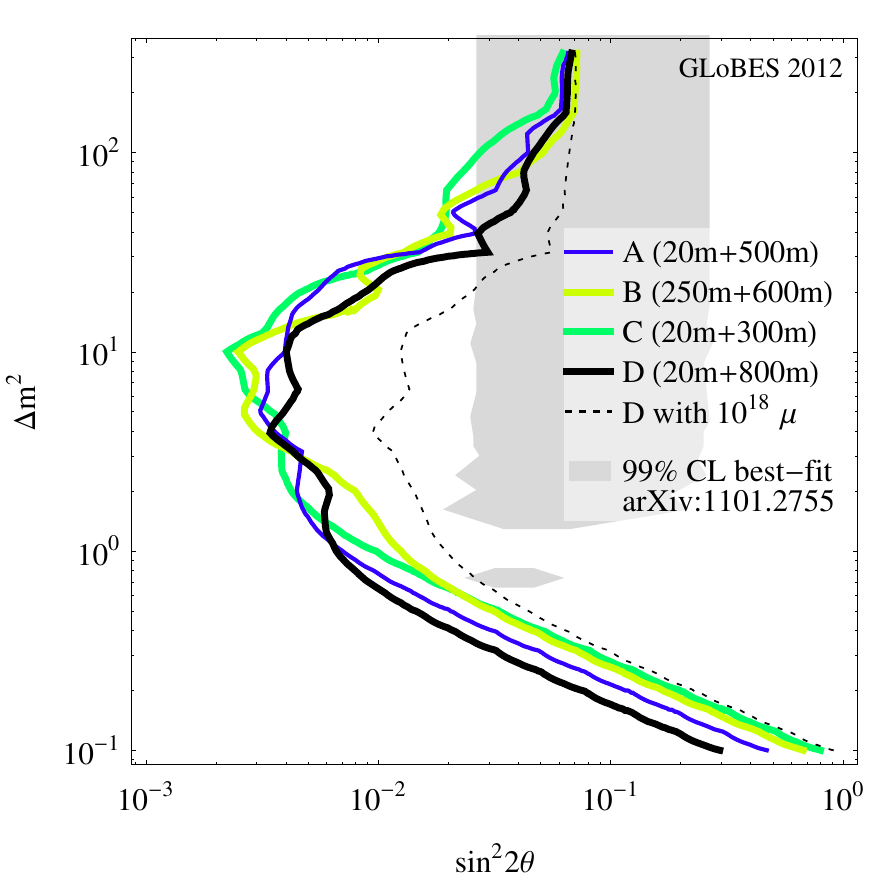} 
\end{center}
\mycaption{\label{fig:opt4} Left panel: Optimal detector distances for $\nu_e$ disappearance, $E_\mu=4 \, \mathrm{GeV}$, $10^{19}$ useful muon decays per polarity. The contours correspond to sensitivities in $\mathrm{log}_{10} (\sin^2 2 \theta)$ down to -2.15 ($\Delta m^2 = 1 \, \mathrm{eV}^2$), -2.5 ($\Delta m^2 = 10 \, \mathrm{eV}^2$), -1.85  ($\Delta m^2 = 32 \, \mathrm{eV}^2$), -1.45 ($\Delta m^2 = 100 \, \mathrm{eV}^2$)  for the depicted values of $\Delta m^2$ (90\% CL, 2 d.o.f.). Right panel: Sensitivities in $\sin^2 2 \theta$-$\Delta m^2$ for the ``optimal'' setups marked in the left panel. For comparison, setup~D is shown with $10^{18}$ useful muon decays per polarity, and the best-fit region from \Ref~\cite{Mention:2011rk} (Fig.~8) is shown as gray-shaded region (99\% CL). }
\end{figure}

In \figu{opt4}, we perform a similar analysis for $E_\mu=4\, \mathrm{GeV}$. In the left panel (two-baseline optimization), only the the point~A is chosen as for $E_\mu=2 \, \mathrm{GeV}$. The qualitative results of the two-baseline optimization are however similar to the above case, with the exception that the optimal region with a long baseline has disappearance in the chosen baseline window. In addition, for very large $\Delta m^2$, somewhat longer baselines are preferred to avoid the geometry effects.  In the right panel, the sensitivities for the chosen points are shown, with rather similar results. Again, point~A appears to be a good compromise between the small and large $\Delta m^2$ sensitivities, but point~D performs better for small $\Delta m^2$. The absolute sensitivities are significantly better than in the $E_\mu=2 \, \mathrm{GeV}$ case, which is a result qualitatively different from the appearance optimization in 
\Ref~\cite{Tunnell:2011ya}, because statistics are important for the disappearance channels. In this case, even the low luminosity curve with 10$^{18}$ useful muon decays covers the relevant parameter space. In summary, for electron neutrino disappearance, either  10$^{19}$ useful muon decays or $E_\mu=4 \, \mathrm{GeV}$ are required to exclude the discussed best-fit parameter space region.

\section{Results and comparison}
\label{sec:results}

\begin{figure}[t!]
\begin{center}
\includegraphics[width=0.48\textwidth]{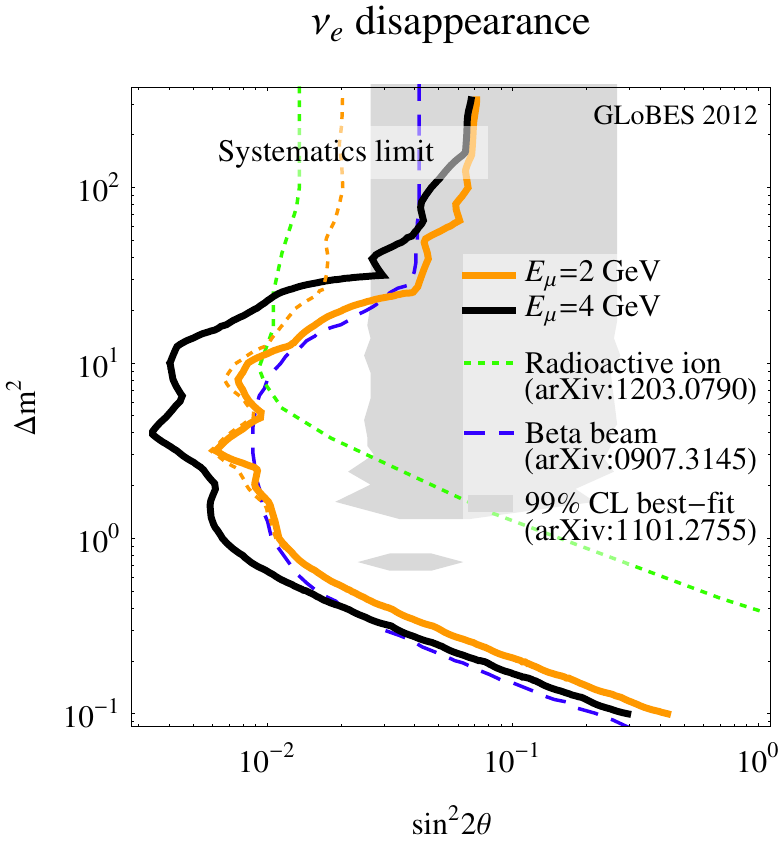} \hspace{0.02\textwidth} \includegraphics[width=0.48\textwidth]{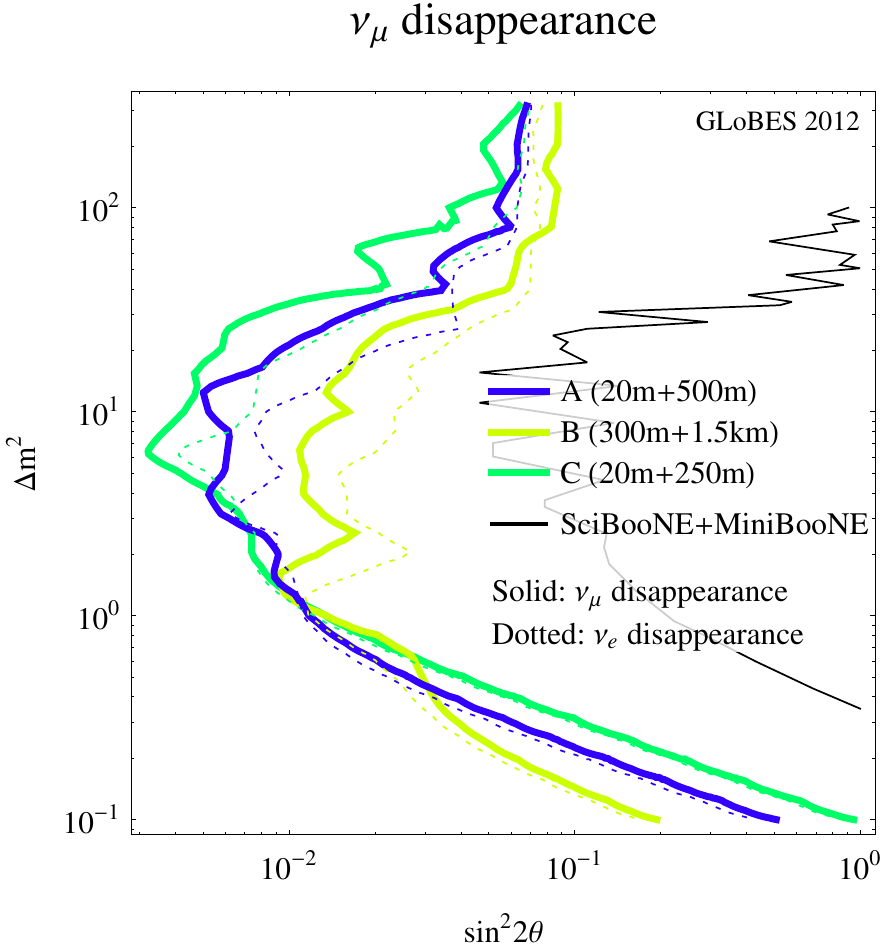}
\end{center}
\mycaption{\label{fig:sens} Left panel: Exclusion region in $\sin^2 2 \theta$-$\Delta m^2$ (right hand sides of curves) for $\nu_e$ disappearance with $10^{19}$ useful muon decays per polarity  for $E_\mu=2 \, \mathrm{GeV}$ (point~A, $20 \, \mathrm{m}+500 \, \mathrm{m}$) and  $E_\mu=4 \, \mathrm{GeV}$ (point~D, $20 \, \mathrm{m}+800 \, \mathrm{m}$) at the 90\% CL, 2 d.o.f. The orange-dashed curve shows the $E_\mu=2 \, \mathrm{GeV}$ result for an improved (2\%) shape error. For comparison, two reference setups are shown: a radioactive ion facility~\cite{Espinoza:2012jr} (dashed-dotted red curve in Fig.~6 therein; $10^{14}$ $^8$Li ions per second) and a low gamma beta beam~\cite{Agarwalla:2009em} (red curve in Figs.~4 and 5 therein). Right panel: Comparison between $\nu_\mu$ disappearance (solid) and $\nu_e$ disappearance (dotted) for the three test points ($10^{19}$ useful muon decays per polarity, $E_\mu=2 \, \mathrm{GeV}$; 90\% CL, 2 d.o.f). The combined SciBooNE and MiniBooNE $\nu_\mu$ disappearance result from \Ref~\cite{Mahn:2011ea} is shown for comparison.}
\end{figure}

We summarize our results in \figu{sens} for $\nu_\mu$ disappearance (left panel) and $\nu_e$ disappearance (right panel). For $\nu_\mu$ disappearance (left panel), optimized setups for different values of $E_\mu$ are compared. As discussed above, the sensitivity for $E_\mu=4 \, \mathrm{GeV}$ is significantly better than for $E_\mu=2 \, \mathrm{GeV}$, but both setups can in principle cover the relevant parameter region. The large $\Delta m^2$ region coverage, marked ``systematics limit'', depends on the assumed external knowledge of cross sections $\times$ efficiencies. The modification of the $E_\mu=2 \, \mathrm{GeV}$ case for an improved 2\% error is shown as dashed curve in the same color. In this case, one can clearly fully cover the discussed (gray-shaded) parameter space. In addition, we show two curves for alternative approaches to $\nu_e$ disappearance measurements. One example is a low $\gamma \simeq 30$ beta beam~\cite{Agarwalla:2009em}, shown as dashed curve. This setup is in a way very similar to ours both in terms of $\gamma$ (our $\gamma \simeq 19$ for $E_\mu=2 \, \mathrm{GeV}$) and the beam geometry. For the detection reaction, however, inverse beta decay is used, and thus the systematical error is assumed to be controllable at the level of 1\% (ours: 10\%). Furthermore, this detection process is only sensitive to $\bar \nu_e$. Another example, shown as dashed curve, is the radioactive ion facility in \Ref~\cite{Espinoza:2012jr}. Here $\gamma \sim 1$ ions are injected into a $4\pi$ detector. Again, the systematical error is assumed to be 1\%. The final result depends on the ion intensity, where we have shown the most aggressive scenario in \figu{sens} ($10^{14}$ ions per second). Here the main issue seems to be the coverage of the low $\Delta m^2$ region, which requires relatively long baselines (which cannot be realized within such a detector). 

For $\nu_\mu$ disappearance (right panel, solid curves), we have used the same fiducial masses $\times$ efficiencies for the sake of simplicity. Therefore, the main difference to $\nu_e$ disappearance are the beam spectrum peaking at higher energies (for the same $E_\mu$), and the better energy resolution. For the optimization, we have not find any qualitative differences compared to $\nu_e$ appearance, apart from the fact that slightly longer baselines are preferred, especially for point~B in \figu{opt2}. For the sake of consistency, we therefore show the same optimization points in \figu{sens} as in \figu{opt2}, and we show the $\nu_e$ disappearance (dotted curves) for comparison. One can easily see that the $\nu_\mu$ disappearance has a slightly better absolute performance, which mainly comes from the higher beam energy. In order to compare the absolute performance to existing experiments, we show the combined SciBooNE and MiniBooNE $\nu_\mu$ disappearance result from \Ref~\cite{Mahn:2011ea}, which was obtained in a similar spirit, as solid thin curve. The VLENF could improve this by about an other of magnitude.

\begin{figure}[t!]
\begin{center}
\includegraphics[width=0.48\textwidth]{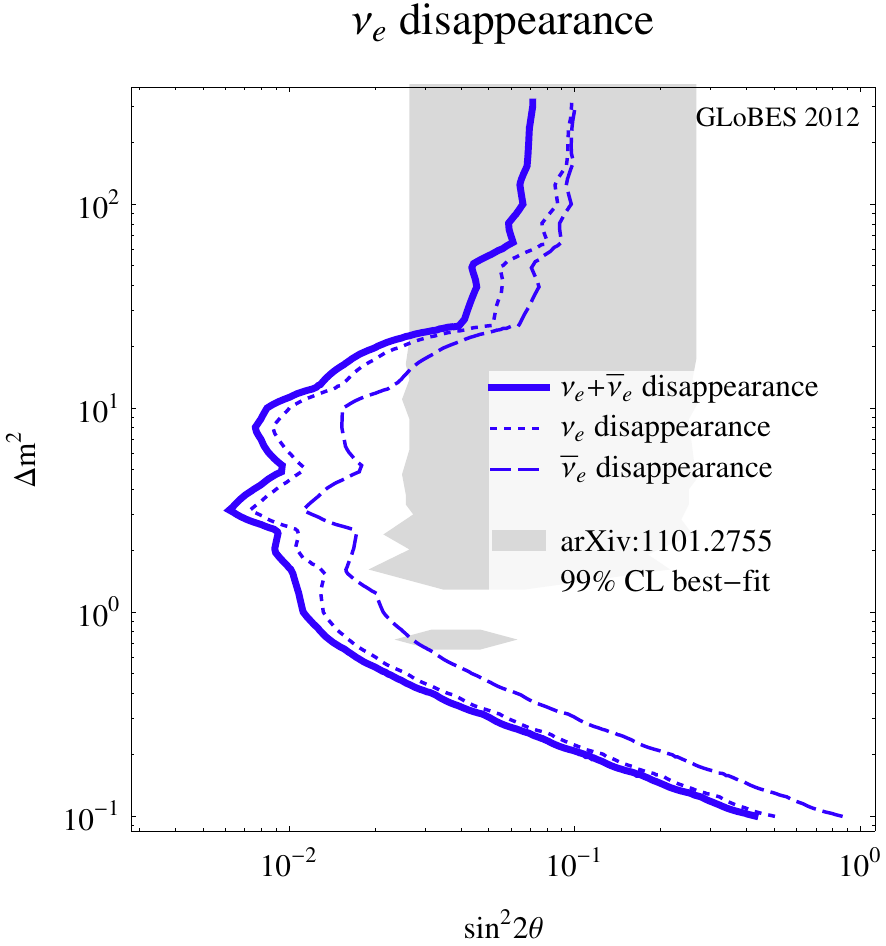} \hspace{0.02\textwidth} \includegraphics[width=0.48\textwidth]{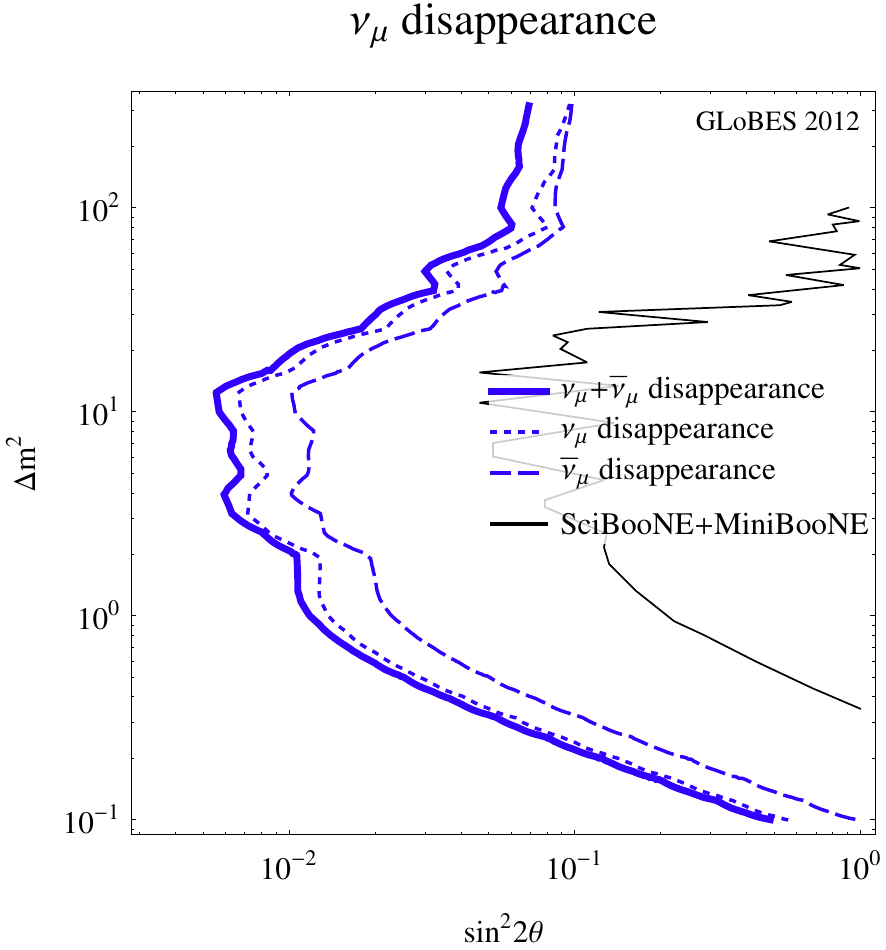}
\end{center}
\mycaption{\label{fig:sep} Contribution of neutrino and antineutrino modes compared to the total result assuming CPT invariance (solid curves) for $\nu_e$ disappearance (left panel) and $\nu_\mu$ disappearance (right panel) at the 90\% CL, 2 d.o.f. .
In both cases, $10^{19}$ useful muon decays per polarity, $E_\mu=2 \, \mathrm{GeV}$, and optimization point~A ( $20 \, \mathrm{m}+500 \, \mathrm{m}$)  have been used.}
\end{figure}

So far we have used the neutrino and antineutrino channels simultaneously, assuming CPT invariance. However, since the earlier mentioned anomalies appear for specific polarities, the separate sensitivities for neutrinos and antineutrino disappearance may be important for some models. We therefore show in \figu{sep} the contributions of the neutrino and antineutrino modes, together with the result assuming CPT invariance. As it can be clearly seen,  similar limits can be obtained for the separated  neutrino and antineutrino channels. The antineutrino sensitivities are somewhat weaker than the neutrino sensitivities due to the smaller cross sections. Nevertheless, CPT invariance (or other consistency) tests can be easily performed, see \Ref~\cite{Giunti:2009en} for details.
This is a significant advantage over many other experiments, which typically have the neutrino channels (beam experiments) or the antineutrino channels (reactor experiments) dominate. 

The discussed alternative setups only represent a limited selection of the ideas for sterile neutrino searches, see Appendix~A of \Ref~\cite{SNAC} for a more complete list.  However, none of the proposed options seems to be able to compete with the proposed disappearance searches at the VLENF.

\section{Summary and conclusions}
\label{sec:summary}

We have studied short-baseline electron and muon neutrino disappearance at a VLENF with $E_\mu=2 \, \mathrm{GeV}$ and $E_\mu=4 \, \mathrm{GeV}$. Compared to the appearance channels, where backgrounds limit the sensitivities, the disappearance channels suffer from the knowledge on cross sections $\times$ efficiencies. We have therefore chose a setup similar to reactor experiments, such as Double Chooz and Daya Bay, using a near and far detector, in which the unknowns can be measured in a self-consistent way. For the systematics, we have adopted the most conservative case of a shape error fully uncorrelated among the bins, but fully correlated between the near and far detectors. In addition, we have included the extension of the decay straight and the beam divergence from the muon decay kinematics, which affect the sensitivity to very large $\Delta m^2 \gtrsim 30 \, \mathrm{eV}^2$ in the near detector. Note that any additional machine-dependent divergence  will add to the muon decay systematics and lead to some additional averaging. However, the muon decay kinematics cannot be eliminated, even if the other systematics might be improved.

We have performed a two-baseline optimization of the setup, where we have identified optimal points depending on the value of $\Delta m^2$. From the different options, we have chosen a setup with the near detector as close as possible to the source and the far detector at a distance between about 500 and 800 meters, which is consistent with the optimization for appearance~\cite{Tunnell:2011ya} and a good compromise between the small and large $\Delta m^2$ sensitivities. As far as the minimal luminosity and muon energy are concerned, we have found that  at least $10^{19}$ useful muon decays per polarity are needed for $E_\mu=2 \, \mathrm{GeV}$, or, alternatively, a higher muon energy, in order to outperform practically any other proposed alternative setup and test the relevant parameter space. That is different from the appearance channel optimization, where lower luminosities may be sufficient and the muon energy hardly matters as long as $E_\mu \gtrsim 2 \, \mathrm{GeV}$~\cite{Tunnell:2011ya}. Note that the VLENF setup can measure  both electron and muon neutrino disappearance, for both neutrinos and antineutrinos. We have also demonstrated that the proposed setup is practically insensitive to the external knowledge on cross sections and efficiencies for $\Delta m^2 \lesssim 30 \, \mathrm{eV}^2$, whereas the sensitivity for larger $\Delta m^2$ depends on the systematics assumptions since oscillations average out in both near and far detectors. 

In conclusion, it is well known from $3+1$ models of sterile neutrinos that disappearance channels provide complementary and important information to constrain the models for sterile neutrinos. In the $3+1$ case, the tension between the appearance and disappearance of various experiments, and between neutrino and antineutrino appearance in MiniBooNE have basically ruled out this model. Apart from the direct tests of disappearance anomalies, for more complicated $3+N$ scenarios, better disappearance information from both  electron and muon neutrino disappearance will be needed. These channels may be uncorrelated in the underlying physics model, as it is evident already in the $3+1$ case. The VLENF can provide this information if it is designed similar to the reactor experiments, with near and far detectors as similar as possible.

\subsubsection*{Acknowledgments}
I would like to thank Alan Bross and Christopher Tunnell for useful discussions, and 
 I would like to acknowledge support from DFG contracts WI 2639/3-1 and WI 2639/4-1


\end{document}